\documentstyle[12pt]{article}
\begin{document}

\begin{center}
{\large COVARIANTLY QUANTIZED  SPINNING PARTICLE AND ITS POSSIBLE CONNECTION TO
NON-COMMUTATIVE
SPACE-TIME }\\
\vskip 2cm
Subir Ghosh\\
\vskip 1cm
Physics and Applied Mathematics Unit,\\
Indian Statistical Institute,\\
203 B. T. Road, Calcutta 700108, \\
India.
\end{center}
\vskip 3cm
{\bf Abstract:}\\
Covariant quantization of the Nambu-Goto spinning particle
in 2+1-dimensions is studied. The model is relevant in the context
of recent activities in non-commutative space-time. From a
technical point of view also covariant quantization of the
model poses an interesting problem: the set of second class
constraints (in the Dirac classification scheme) is {\it
 reducible}. The reducibility problem  is analyzed from two contrasting
 approaches: (i) the auxiliary variable method \cite{bn}  and (ii) the
 projection operator method \cite{blm}. Finally in the former
 scheme, a Batalin-Tyutin quantization has been done. This
 induces a mapping between the non-commutative and the ordinary space-time. BRST quantization programme in the latter scheme has also been discussed.


\newpage
The recent activity in Non-Commutative (NC) field theory \cite{dn}
and more generally
concerning NC space-time \cite{nc} has recreated a lot of interest in
the study of
physically motivated models, where the NC feature appears naturally. The
 well-known
Landau problem, (of a charged particle confined in a plane in the presence
of a magnetic
field in the perpendicular direction), is one such example. In an earlier
 paper \cite{sg1}
we have pointed out that  the bosonic Spinning Particle Model (SPM),
originally proposed
by Hanson and Regge \cite{hr}, is relevant for theories in
NC space-time. As we
shall show in detail, (this was also pointed out briefly in \cite{sg1}),
the SPM can
provide a direct mapping between ordinary (that is commuting) and NC
space-time.

The Nambu Goto construction of the SPM \cite{hr}, by itself, is an
interesting example
of a relativistic theory having a non-linear and non-abelian constraint
structure. An
added feature is that the  system of
Second Class Constraints \cite{d} is {\it reducible} in nature if
 manifest
Lorentz covariance is to be maintained. The present work focuses on the
last point since
some of the corresponding  results in a non-covariant setup have already
been presented by
us in \cite{sg1}. Quite obviously the non-covariant results are somewhat
inelegant and
will be difficult to use in a relativistic theory. We will discuss the
preliminary steps leading towards a
BRST quantization of the SPM in a manifestly covariant way, along the lines of
\cite{bfv,bt}.

According to the Dirac classification scheme of Hamiltonian analysis of a
constraint
system \cite{d}, the SPM has both First Class Constraints (FCC) and a
reducible
system of Second Class Constraints (SCC). The former generate gauge
invariance whereas the
latter restrict the phase space manifold along with a modification in the
canonical symplectic structure.
Reducibility in a non-linear SCC system is a novel feature and
possibly the present work is the first
example where
a nonlinear reducible SCC system is quantized. The reducibility problem
for the SCCs
of the SPM in a covariant framework
has to be addressed first before one can embark upon a conventional
BRST quantization of a set of reducible FCCs. The problem of reducibility
will be handled by
 two very distinct approaches, {\it i.e.} the Auxiliary
Variable method
\cite {bn} and the Projection Operator method \cite{blm}.

The BRST programme \cite{bfv,bt}, in Auxiliary Variable method \cite{bn},
proceeds in three stages: In stage (i) the reducibility in the SCC system is
taken care of by introducing a set of auxiliary degrees of freedom. This
enlargement of the phase
space modifies the original reducible
SCCs and converts
them in to an {\it irreducible} (or independent) set.
 However, care should be taken to see that the
{\it extension does
 not affect the physical, (i.e. original), phase space}. In stage (ii),
the set of irreducible SCCs are further modified by bringing in
the Batalin-Tyutin \cite{bt} variables so that the SCCs are
transformed to FCCs. Subsequently in stage (iii), the conventional BRST
 quantization is to be performed. Note that
  no ghost for ghosts appear here, (as is customary
in any reducible theory), since the reducibility is removed in stage (i).

In the Projection Operator formalism \cite{blm},
the reducibility problem is solved by the
 construction of a projection operator
which projects
out a
maximal set of weakly involutive constraints, {\it i.e.}
FCCs,  (from the SCC system), with which a generalization of the
standard BRST
quantization is possible. In this scheme,  ghost for ghosts do appear. It might
be interesting to see
if the auxiliary variables of the former method and secondary ghosts of the latter
method are related.
Indeed, the inherent nonlinearity
and non-abelian nature of the SPM
constraint system is a real test of the viability of
the above
schemes \cite{bn,blm} in arbitrary models.

The connection between SPM and NC space-time  is discussed here in
the auxiliary variable method. After the second stage of extension
of the phase space, (where the BT variables appear), we
demonstrate the existence of a mapping between the NC space-time
coordinate and the normal one, via the auxiliary BT degrees of
freedom \cite{bt}. The analogous results in a manifestly
non-covariant setup were derived in \cite{sg1}.
 Following the BT prescription \cite{bt} in
the SPM, the NC space-time coordinate operators are expressible as
normal space-time coordinates, appropriately extended by  BT
contributions. This provides the mapping between the NC and
ordinary space-times in  the extended phase space. As has been
noticed in earlier
 studies of nonlinear theories
\cite{bbg,sg1}, the possibility  of an infinite number of higher order
Batalin-Tyutin (BT) variable contributions in some of the physically relevant operators
manifests here also.

The paper is organized as follows. In section {\bf II} a brief
resume$'$ of the SPM is provided, which also helps us to fix the
notations. Section {\bf III} deals with the application of the
Auxiliary Variable method in SPM. In section {\bf IV}, the relevant
formulas for the BT quantization are provided and subsequently the
method is applied to the irreducible set of SCCs in SPM, obtained
in the previous section. The connection with the NC space-time is
also elucidated here.  In section {\bf V} the Projection
Operator in the context of SPM is derived.  Necessary steps for the subsequent BRST quantization in the present case has been provided. Determination of
the explicit structure of the Projection operator in a complex
model is very important since its existence was only suggested in
\cite{blm}. Sections {\bf III}-{\bf V} comprise
the main body of the work.
The paper is concluded with a discussion in section
{\bf VI}.

\begin{center}
{\bf II. SPINNING PARTICLE MODEL: A BRIEF RESUME$'$}
\end{center}
The 3+1-dimensional Nambu-Goto Lagrangian of the SPM,
originally proposed by Hanson and Regge \cite{hr} is,
\begin{equation}
L=[M^2a_1+{{J^2}\over 2}a_2+2MJ({1\over 2}a_1a_2+a_3)^{{1\over
2}}]^{{1\over 2}}.
 \label{1}
\end{equation}
The notations of \cite{hr} are used, where
$$u^\mu={{dx^\mu }\over {d\tau }} ~~,~~\sigma ^{\mu\nu }=
\Lambda _\rho ^{~\mu } {{d\Lambda ^{\rho\nu }}\over {d\tau
}}=-\sigma ^{\nu\mu }$$
and the dynamical variables entering $L$ are
\begin{equation}
a_1=u^\mu u_\mu ~,~a_2=\sigma ^{\mu\nu }\sigma _{\mu\nu }
~,~a_3=u_\mu \sigma ^{\mu\nu }\sigma _{\nu \lambda } u^\lambda
~,~a_4=det \sigma ={1\over {16}}(\sigma ^{\mu\nu }\sigma ^*
_{\mu\nu })^2 ~,~\sigma ^*_{\mu\nu }={1\over 2}\epsilon
_{\mu\nu\rho\lambda }\sigma ^{\rho\lambda }.
\label{2}
\end{equation}
Here $(x^\mu ~,~\Lambda ^{\mu\nu })$ is a Poincare group element
and also a set of dynamical variables of the theory, with
$$
\Lambda _\rho ^{~\mu }\Lambda^{\rho\nu } =\Lambda _{~\rho
}^\mu \Lambda^{\nu \rho }=g^{\mu\nu}~~,~~g^{00}=-g^{ii}=1.
$$
In order to discuss the Hamiltonian formulation, we define the canonically
conjugate momenta as,
\begin{equation}
 P^\mu\equiv {{\partial L}\over {\partial
u_\mu}}
=L^{-1}[M^2u^\mu +{{MJ}\over {({{a_1a_2}\over
2}+a_3)}^{1\over 2}}({1\over 2}a_2u^\mu+\sigma ^{\mu\nu }\sigma
_{\nu \lambda } u^\lambda )],
\label{3}
\end{equation}
\begin{equation}
S^{\mu\nu}\equiv {{\partial L}\over {\partial \sigma _{\mu\nu
}}}=L^{-1}[J^2\sigma ^{\mu\nu } +{{MJ}\over {({{a_1a_2}\over
2}+a_3})^{1\over 2}}(a_1\sigma ^{\mu\nu }+(u^\mu\sigma
^{\nu\lambda}- u^\nu\sigma ^{\mu\lambda})u_\lambda )].
 \label{4}
\end{equation}
One immediately finds the primary constraints,
\begin{equation}
P^\mu P_\mu =M^2(1+{{J^2}\over{L^2}}{{a_1a_4}\over {({{a_1a_2}\over
2}+a_3)}}) ~~,~~S^{\mu\nu} S_{\mu\nu}=2J^2 ~~,~~S^{\mu\nu}P_\nu =0.
\label{5}
\end{equation}
This particle model is somewhat unconventional because of the operator valued
"mass" which can only reduce to the standard form for $a_4\approx 0$.
However, we have
shown \cite{sg2,sg1} that in 2+1-dimensions, this complication can be avoided with
the Lagrangian posited by us,
\begin{equation}
L=[M^2a_1+{{J^2}\over 2}a_2+MJ\epsilon ^{\mu\nu\lambda}u_\mu
\sigma _{\nu\lambda } ]^{{1\over 2}} .
\label{6}
\end{equation}
With the conjugate momenta,
\begin{equation}
P^\mu={{\partial L}\over {\partial u_\mu}} =L^{-1}[M^2u^\mu
+{{MJ}\over 2}\epsilon ^{\mu\nu\lambda}\sigma _{\nu\lambda }]
\label{7}
\end{equation}
\begin{equation}
S^{\mu\nu}={{\partial L}\over {\partial \sigma _{\mu\nu
}}}={{L^{-1}}\over 2}[J^2\sigma ^{\mu\nu } +{MJ}\epsilon
^{\mu\nu\lambda}u_\lambda ]
 \label{8}
\end{equation}
we obtain the constraints as,
\begin{equation}
P^\mu P_\mu =M^2 ~~,~~S^{\mu\nu} S_{\mu\nu}=2J^2,
\label{9}
\end{equation}
\begin{equation}
S^{\mu\nu}P_\nu =0.
\label{10}
\end{equation}
(\ref{9}) constitutes the Casimir operators. This model has been
successfully used \cite{sg2,sg1}
 in the context of {\it anyons}, {\it i.e.}
excitations in 2+1-dimensions, having arbitrary spin and statistics
\cite{fw}. Since the NC feature of the resulting space-time coordinates
is also preserved in 2+1-dimensions from now on we will work in 2+1-dimensions. An additional
set of constraints are introduced, (for a detailed discussion see \cite{hr,sg2})
and the full set of constraints are,
\begin{equation}
\Psi _1\equiv P^\mu P_\mu -M^2 ~~,~~\Psi _2\equiv S^{\mu\nu} S_{\mu\nu}-2J^2,
 \label{11}
\end{equation}
\begin{equation}
\Theta _1^\mu \equiv S^{\mu\nu}P_\nu~~~,~~~\Theta _2 ^\mu \equiv \Lambda
^{0\mu}-{{P^\mu }\over M}~~,~~\mu =0,~1,~2~.
 \label{12}
\end{equation}
\footnote {Note that instead of $\Psi _2$ as above, one can equivalently use
$\Psi _2\equiv \epsilon
^{\mu\nu\lambda}S_{\mu\nu}P_\lambda -MJ, $
which incidentally defines the Pauli Lubanski scalar.}
With the help
of the following canonical Poisson Brackets (PB),
\begin{equation}
\{P^\mu ,x^\nu \}=g^{\mu\nu}~~,~~\{P^\mu ,P^\nu \}=0 ~~,~~\{x^\mu
,x^\nu \}=0
 \label{011}
\end{equation}
$$
\{S^{\mu\nu},S^{\lambda\sigma}\}=S^{\mu\lambda}g^{\nu\sigma}-S^{\mu\sigma}g^{\nu\lambda}
+S^{\nu\sigma}g^{\mu\lambda}-S^{\nu\lambda}g^{\mu\sigma},$$
\begin{equation}
\{\Lambda
^{0\mu},S^{\nu\sigma}\}=\Lambda ^{0\nu}g^{\mu\sigma}-\Lambda
^{0\sigma}g^{\mu\nu}~~,~~\{\Lambda^{0\mu},\Lambda ^{0\nu}\}=0
\label{012}
\end{equation}
we compute the constraint algebra, where $\Psi _1$ trivially
commutes with all the constraints and the rest of the non-zero
PBs between the constraints are,
\begin{equation}
\{\Psi _2,\Theta _2^\mu\}=4(S^{\mu\lambda}\Theta _{2 \lambda}
+{1\over
M}\Theta _1^\mu )~~,~~\{\Psi _2,\Theta _1^\mu\}=0 \label{13}
\end{equation}
\begin{equation}
\{\Theta ^\mu _\alpha  ,\Theta ^\nu _\beta \}\equiv\Delta ^{\mu\nu}_
{\alpha\beta}~~,~~\alpha ,\beta =1,2~~,
\label{14}
\end{equation}
where,
$$
\Delta ^{\mu\nu}_{12} \equiv
\{\Theta _1^\mu ,\Theta _2^\nu\}=\frac{1}{M}(M^2g^{\mu\nu}-P^\mu P^\nu)
-P^\nu
\Theta _2^\mu +g^{\mu\nu}(P.\Theta _2)+\frac{1}{M}g^{\mu\nu}\Psi_1
~, $$
\begin{equation}
\Delta ^{\mu\nu}_{22}\equiv
\{\Theta _2^\mu ,\Theta _2^\nu\}=0~ ,~
\Delta ^{\mu\nu}_{11}\equiv
\{\Theta _1^\mu ,\Theta _1^\nu\}=M^2S^{\mu\nu}+\Psi _1S^{\mu\nu}
+P^\mu \Theta _1^\nu
-P^\nu \Theta _1^\mu . \label{014}
\end{equation}
One can see that the constraint algebra for $\Psi _\alpha
,~(\alpha =1,2)$, closes whereas, $\Delta ^{\mu\nu}_{\alpha\beta}$
being non-trivial even on the constraint surface,
 indicates the presence of SCCs. Hence, according
to the Dirac classification scheme \cite{d}, $\Psi _\alpha $ and
$\Theta _\alpha ^\mu $ constitute FCCs and SCCs respectively.
Demanding time persistence of the FCCs will generate no further
constraints since the theory being reparametrization invariant,
its Hamiltonian will be a combination of FCCs only.

It is not possible to compute the Dirac Brackets (DB) \cite{d} from
the SCCs since the system of SCC is reducible ({\it i.e.} not
independent) due to the following identity,
\begin{equation}
P_\mu \Theta _1^\mu =0.
\label{15}
\end{equation}
Note also the presence of the relation,
\begin{equation}
(\Lambda ^0_\mu +\frac{P_\mu}{M})\Theta _2^\mu =-\frac{\Psi_1}{M^2}.
\label{015}
\end{equation}
However, since (\ref{015}) involves an FCC,
this is not a reducibility condition \cite{ht} and only
restricts the number of independent degrees of freedom
on the constraint manifold. Also, this system is first stage reducible
since higher order reducibility conditions are absent.

The Hamiltonian of the system turns out to be that of a free particle due to the so called "rigidity" property of the particle \cite{cnp,sg2}  meaning that when the SCCs are enforced strongly, the spin vector becomes proportional to the momentum vector. As stated before, the Hamiltonian being a combination of the FCCs (\ref{9}), due to the above reason, it is sufficient to consider only the mass shell condition  $\Psi _1$ in (\ref{9}). One has to fix the time scale by choosing a gauge for the FCC $\Psi _1$, which can simply be
$$x_0=\tau ~.$$
Subsequently one has to construct the DBs for the above SCC pair and the Hamiltonian is obtained from $\Psi _1=0$,
\begin{equation}
H\equiv P_0= \sqrt{P_iP_i+M^2}.
\label{ham}
\end{equation}

Indeed, one can obtain the
DBs by considering an irreducible set from $\Theta _\alpha ^\mu$, ({\it e.g.}
taking only the spatial components $\Theta _\alpha ^i$, as in \cite{sg1}),
 but
this destroys the manifest covariance of the model. We now follow the method prescribed by Banerjee and
Neto \cite{bn} to obtain the DBs without losing manifest covariance.
\vskip .2cm
\begin{center}
{\bf III. COVARIANT QUANTIZATION:  AUXILIARY FIELD METHOD}
\end{center}
\vskip .1cm In the formalism proposed in \cite{bn}, the reducible
SCCs ($\Theta _\alpha ^\mu $) are modified in an appropriate way
by introducing auxiliary degrees of freedom, such that the
modified SCCs ($\bar\Theta _\alpha ^\mu $) become irreducible. At
the same time, one has to ensure that the extension does not
affect the physical phase space. In a practical sense, this means
that the resulting DBs between the physical ({\it i.e.} original)
degrees of freedom will have to be independent of the auxiliary
variables or any parameters connected to them. The reducibility
condition plays a crucial role in determining the structures of
the modification terms (in the SCCs), which have to be such that
on imposition of the reducibility conditions on the SCCs, the
auxiliary variables vanish exactly.

The phase space is extended by introducing a canonical pair of auxiliary
variables $\phi$ and $\pi$ that satisfy $\{\phi ,\pi \}=1$ and PB commute
with the rest of the physical variables. The SCCs $\Theta _\alpha ^\mu $
are modified in the way as shown below,
\begin{equation}
\bar \Theta _1^\mu\equiv S^{\mu\nu}P_\nu +k_1P^\mu \pi ~~~;
~~~\bar \Theta _2^\mu
\equiv (\Lambda ^{0\mu}-\frac {P^\mu}{M})+k_2 (\Lambda
^{0\mu}+\frac {P^\mu}{M})\phi ~, \label{150}
\end{equation}
where $k_1$ and $k_2$ denote two arbitrary parameters.
The constraint matrix is computed to be,
\begin{equation}
\bar\Delta _{\alpha\beta}^{\mu\nu}\equiv\{\bar\Theta _\alpha ^\mu ,
\bar\Theta _\beta ^\nu\}=
 \left (
\begin{array}{cc}
 \bar\Delta _{11} ^{\mu\nu} &  \bar\Delta _{12} ^{\mu\nu}\\
-\bar\Delta _{12} ^{\nu\mu} &  \bar\Delta _{22}^{\mu\nu}
\end{array}
\right ) \label{18}
\end{equation}
where,
$$
\bar \Delta _{11} ^{\mu\nu} \equiv
\{\bar\Theta _1 ^\mu ,
\bar\Theta _1 ^\nu\}= \Delta _{11} ^{\mu\nu}  =
M^2S^{\mu\nu}+\Psi _1S^{\mu\nu}
+P^\mu S^{\nu\lambda}P_\lambda
-P^\nu  S^{\mu\lambda}P_\lambda ,
$$

$$
\bar \Delta _{21} ^{\mu\nu}
\equiv\{\bar\Theta _2 ^\mu ,
\bar\Theta _1^\nu\}=
\frac{1}{M} [r(1+k_2\phi
)+k_1k_2(1+r)]P^\mu P^\nu-Mr(1+k_2\phi)g^{\mu\nu},
$$

$$
\equiv r_1P^\mu P^\nu +r_2g^{\mu\nu} ,
$$

\begin{equation}
\bar \Delta _{22} ^{\mu\nu}
\equiv\{\bar\Theta _2 ^\mu ,
\bar\Theta _2^\nu\}= 0,
\label{19}
\end{equation}
with $r_1$ and $r_2$ given by,
$$r_1=\frac{1-k_2^2\phi ^2+2k_1k_2}{M(1+k_2\phi )} ~~,~~r_2=-M(1-k_2\phi ).$$
In the above, we have used
\begin{equation}
\Lambda^{0\mu}=-\frac{(k_2\phi -1)P^\mu}{(k_2\phi +1)M}\equiv
\frac{r}{M}P^\mu ~,\label{16}
\end{equation}
which follows directly from $\bar\Theta _2^\mu =0$.

Let the inverse of $\bar\Delta _{\alpha\beta}^{\mu\nu} $ be defined as,
\begin{equation}
\bar\Delta ^{\mu\nu}_{\alpha \beta }\bar\Delta _{\nu\lambda}^{\beta \gamma }
=\delta ^\mu_\lambda
\delta _\alpha ^\gamma ,
\label{inv}
\end{equation}
and we consider a general form of the inverse matrix to be,
\begin{equation}
\bar\Delta _{\nu\lambda}^{\alpha \beta } =
\left (
\begin{array}{cc}
 0 & -a_{\nu\lambda}\\
a_{\nu\lambda} & b_{\nu\lambda}
\end{array}
\right )
\label{019}
\end{equation}
with the entries,
\begin{equation}
a_{\nu\lambda}\equiv a_1P_\nu P_\lambda +a_2g_{\nu\lambda}~~,~~
b_{\nu\lambda}\equiv b(\bar \Delta _{11})_{\nu\lambda} .
\label{20}
\end{equation}
The parameters $a_1,~a_2$ and $b$ are found to be,
\begin{equation}
a_1=-\frac{1-k_2^2\phi ^2+2k_1k_2}{2M^3k_1k_2(1-k_2\phi )}~~,~~a_2=-\frac{1}{r_2}
~~,~~b=\frac{1}{r_2^2}.
\label{21}
\end{equation}
The
DBs \cite{d} are now calculated
for any two generic variables $A$ and $B$ from the defining equation,
\begin{equation}
\{A,B\}_{DB}=\{A,B\}-\{A,\bar\Theta _\alpha ^\mu\}
\bar\Delta _{\mu\nu}^{\alpha \beta }\{\bar\Theta _\beta ^\nu, B\} ,
\label{22}
\end{equation}
where $\bar\Delta _{\nu\lambda}^{\alpha \beta }$ is defined in
(\ref{inv}).

After a long and quite involved algebra we recover the DBs of the
original (physical) variables in
a covariant form,
\begin{equation}
\{x^\mu ,x^\nu\}_{DB}=-\frac{S^{\mu\nu}}{M^2}~,~\{P^\mu ,x^\nu\}_{DB}=
g^{\mu\nu}~,~
\{P^\mu ,P^\nu\}_{DB}=
\{P^\mu ,\Lambda^{\nu\lambda}\}_{DB}=\{P^\mu ,S^{\nu\lambda}\}_{DB}=
0~,
\label{xx}
\end{equation}
$$
\{x^\mu ,\Lambda^{\sigma\nu}\}_{DB}=\frac{1}{M^2}\Lambda ^{\sigma\alpha }(g^\mu _\alpha
P^\nu -g^{\mu\nu}P_\alpha )~,
$$
\begin{equation}
\{x^\mu ,S^{\sigma\nu}\}_{DB}=-\frac{1}{M^2}(S^{\mu\sigma}P^\nu-
S^{\mu\nu}P^\sigma ) ~,
\label{x}
\end{equation}

$$\{S^{\mu\nu},S^{\lambda\sigma}\}_{DB}=S^{\mu\lambda}(g^{\nu\sigma}
-\frac{P^\nu P^\sigma }{M^2})
-S^{\mu\sigma}(g^{\nu\lambda}- \frac{P^\nu P^\lambda }{M^2})
+S^{\nu\sigma}(g^{\mu\lambda}-\frac{P^\nu P^\lambda }{M^2})-S^{\nu\lambda}
(g^{\mu\sigma}
-\frac{P^\mu P^\sigma }{M^2}),$$
$$\{\Lambda
^{\alpha\mu},S^{\nu\sigma}\}_{DB}=\Lambda ^{\alpha\nu}(g^{\mu\sigma}-
\frac{P^\mu P^\sigma }{M^2})
-\Lambda ^{\alpha\sigma}(g^{\mu\nu}-
\frac{P^\nu P^\nu }{M^2})+\frac{\Lambda ^{\alpha \beta} P_\beta}
{M^2}(P^\sigma g^{\mu\nu}
-P^\nu g^{\mu\sigma})$$
\begin{equation}
\{\Lambda^{\alpha\mu},\Lambda ^{\beta\nu}\}=0 .
\label{ss}
\end{equation}
It is very important to note that the
{\it DBs between the physical degrees of freedom are totally independent
of  $k_1$ and $k_2$}, the
parameters that appeared
in conjunction with the auxiliary variables $\phi$ and $\pi$. There is
no need to take a vanishing limit of  $k_1$ and $k_2$. This feature
ensures that the phase space extension has not disturbed the
sector of physical variables. The importance of this has been
repeatedly stressed in \cite{bn}.

Notice that the DBs involving $P^\mu$ remain unchanged from the
PBs but the {\it non-commutativity in configuration space is reflected
in the non-zero} $\{x^\mu ,x^\nu\}_{DB}$.

Quite naturally, the DBs constituting an auxiliary variable, such
as the one given below,
\begin{equation}
\{x^\mu,\phi \}_{DB}=\frac{(1-k_2\phi )(1+k_2\phi )}{2k_2M^2} ,
\label{23}
\end{equation}
will involve $k_1$ or $k_2$. Moreover, they will be undefined
for the zero limit of these parameters.
This completes the first stage extension and with this irreducible
as well as covariant set of
SCCs (\ref{150}) we now proceed to the second stage extension in the
Batalin-Tyutin  formalism \cite{bt}.
\vskip .2cm
\begin{center}
{\bf IV: BATALIN-TYUTIN EXTENSION AND NON-COMMUTATIVE SPACE-TIME }
\end{center}
\vskip .1cm
The basic idea behind
the BT scheme \cite{bt,bfv} is to
 introduce additional phase space variables (BT variables) $\phi ^\alpha _a $,
 besides the existing degrees of freedom, such that {\it all}
 the constraints in the extended system are converted to FCCs.
 The advantage is that the FCC system, being a gauge theory, enjoys more
 freedom in the form of choice of gauge in quantization and the quantization
 procedure itself is well understood for a gauge theory in a canonical phase
 space.
 This means
that one has to modify the original constraints and Hamiltonian
accordingly by putting BT-extension terms in them. The way to
achieve this at the classical level has been provided in
\cite{bt}. The main results of the BT prescription \cite{bt}
relevant for our purpose are listed below.

Let us consider a set of constraints $(\bar\Theta _\alpha ^\mu,\Psi _l
)$ and a Hamiltonian operator $H$ with the following PB relations,
$$
\{\bar\Theta ^\mu_\alpha (q) ,\bar\Theta ^\nu_\beta (q)\}\approx
\bar\Delta ^{\mu\nu}_{\alpha \beta
 }(q) \ne 0~~;~~\{\bar\Theta ^\mu_\alpha (q) ,\Psi _l (q)\}\approx 0
$$
\begin{equation}
 \{\Psi _l(q) ,\Psi _n (q)\}\approx 0 ~~;~~
\{\Psi _l(q) ,H (q)\}\approx 0.
\label{01}
\end{equation}
In the above $(q)$
collectively refers to the set of variables
present prior to the BT extension
and "$\approx $"
means that the equality holds on the constraint surface. Clearly
$\bar\Theta
_\alpha ^\mu $
and $\Psi _l $ are SCC and FCC \cite{d} respectively.
These constraints are actually the ones we have been working with,
 {\it i.e.}
$\bar\Theta
_\alpha ^\mu $  are given in (\ref{150}) and  $\Psi _1 $ of the starting
FCCs (\ref{9}) remains unchanged, whereas $\Psi _2$ can be modified to make it
an FCC, (at least up to low order in the auxiliary variables). However, this
restriction is not important for our present discussion.

In systems with non-linear SCCs,
(such as the present one), in general the DBs can become
dynamical variable dependent \cite{bbg,sg1}
due to the $\{A,\bar\Theta _\alpha ^\mu\}$ and
$\bar\Delta ^{\mu\nu}_{\alpha \beta }$
 terms, leading to problems for
the quantization programme. To cure this type of pathology, BT
formalism is a systematic framework where one introduces the BT
variables $\phi
 ^\alpha _a $, obeying
\begin{equation}
\{\phi ^\alpha _\mu,\phi ^\beta _\nu\}=\omega ^{\alpha \beta}_{\mu\nu}=
-\omega ^{\beta \alpha}_{\nu\mu},
\label{bt}
\end{equation}
where $\omega ^{\alpha \beta}_{\mu\nu}$ is a constant (or at most
 a c-number function) matrix, with the aim of modifying the SCC
$\bar\Theta _\alpha ^\mu (q)$ to $\tilde \Theta _\alpha ^\mu (q,
\phi ^\alpha _\mu )$
such that,
\begin{equation}
\{\tilde\Theta ^\mu_\alpha (q,\phi ) ,\tilde\Theta ^\nu_\beta (q,\phi )\}=0
~~;~~\tilde\Theta ^\mu_\alpha (q,\phi )=\Theta ^\mu_\alpha (q)+
\Sigma _{n=1}^\infty \tilde\Theta ^{\mu(n)} _\alpha (q,\phi )~~;~~
\tilde\Theta ^{\mu(n)}\approx O(\phi ^n)
\label{b1}
\end{equation}
This means that $\tilde\Theta ^\mu _\alpha $ are now FCCs and in
particular abelian \cite{bt}. A simple choice, obviously not unique, is
\begin{equation}
\omega ^{\alpha \beta}_{\mu\nu}=g_{\mu\nu}\epsilon^{\alpha\beta}~,~\epsilon^{12}=1.
\label{new}
\end{equation}
The explicit terms in the above expansion are \cite{bt},
\begin{equation}
\tilde\Theta ^{\mu(1)}_\alpha =X^{\mu\nu}_{\alpha \beta }\phi ^\beta _\nu~~;~~
\bar\Delta ^{\mu\nu}_{\alpha \beta }+X^{\mu\lambda}_{\alpha \delta }
\omega _{\lambda\sigma}^{ \delta \gamma }X^{\nu\sigma}_{\beta \gamma }=0
\label{b2}
\end{equation}

\begin{equation}
\tilde\Theta ^{\mu(n+1)}_\alpha =-{1\over{n+2}}
\phi_{\sigma}^{\delta }\omega ^{\sigma\lambda}_{\delta \gamma }X_
{\lambda\nu}^{\gamma \beta }B^{\nu\mu(n)}_{\beta \alpha }~~;~~n\ge 1
\label{b3}
\end{equation}

\begin{equation}
B^{\nu\mu(1)}_{\beta \alpha }=
\{\tilde\Theta ^{\nu(0)} _\beta ,\tilde\Theta ^{\mu(1)} _\alpha \}_{(q)}-
\{\tilde\Theta ^{\mu(0)} _\alpha ,\tilde\Theta ^{\nu(1)} _\beta \}_{(q)}
\label{b4}
\end{equation}

\begin{equation}
B^{\mu\nu(n)}_{\beta \alpha }=
\Sigma _{m=0}^n
\{\tilde\Theta ^{\nu(n-m)} _\beta ,\tilde\Theta ^{\mu(m)} _\alpha \}_{(q,p)}+
\Sigma _{m=0}^n
\{\tilde\Theta ^{\nu(n-m)} _\beta ,\tilde\Theta ^{\mu(m+2)} _\alpha \}_{(\phi )}
~~;~~n\ge 2
\label{b5}
\end{equation}
In the above, we have defined,
\begin{equation}
X^{\mu\nu}_{\alpha \gamma }X_{\nu\lambda}^{\gamma\beta  }=
\omega ^{\mu\nu}_{\alpha \gamma }\omega _{\nu\lambda}^{\gamma\beta  }
=\delta ^\beta _\alpha \delta ^\mu _\lambda .
\label{b6}
\end{equation}
A very useful
idea is to introduce the Improved Variable $\tilde f(q)$ \cite{bt}
 corresponding to each $f(q)$,
\begin{equation}
\tilde f(q,\phi )\equiv f(\tilde q)
=f(q)+\Sigma _{n=1}^\infty \tilde f(q,\phi )^{(n)}~~
;~~\tilde f^{(1)}=-
\phi_{\mu}^{\alpha }\omega ^{\mu\nu}_{\alpha\beta  }X_{\nu\lambda}^{\beta \delta }\{
\bar\Theta_\delta ^\lambda ,f\}_{(q)}
\label{b7}
\end{equation}

\begin{equation}
\tilde f^{(n+1)}=-{1\over{n+1}}
\phi_{\mu}^{\alpha }\omega ^{\mu\nu}_{\alpha\beta }X_{\nu\lambda}
^{\beta \delta }
G(f)^{\lambda (n)}_\delta ~~;~~n\ge 1
\label{b8}
\end{equation}

\begin{equation}
G(f)^{\mu(n)}_{\beta }=
\Sigma _{m=0}^n
\{\tilde\Theta ^{\mu (n-m)} _\beta ,\tilde f^{(m)}\}_{(q)}+
\Sigma _{m=0}^{(n-2)}
\{\tilde\Theta ^{\mu (n-m)} _\beta ,\tilde f^{(m+2)}\}_{(\phi )}
+\{\tilde\Theta ^{\mu (n+1)} _\beta ,\tilde f^{(1)}\}_{(\phi )}
\label{b9}
\end{equation}
which have the property
$\{\tilde\Theta ^\mu_\alpha (q,\phi ) ,\tilde f(q,\phi )\}=0$.
Thus the improved variables are FC or equivalently gauge invariant.
The subscript $(\phi )$ and $(q)$ in the PBs indicate the variables
with respect to which the PBs are to be taken.
It can be
proved that extensions of the original FCC $\Psi _l $ and Hamiltonian
 $H$ are simply,
\begin{equation}
\tilde \Psi _l=\Psi (\tilde q)~~;~~
\tilde H=H (\tilde q).
\label{b10}
\end{equation}
One can also reexpress the converted SCCs as
$\tilde\Theta ^\mu_\alpha \equiv \Theta ^\mu_\alpha (\tilde q)$.
The following identification theorem,
\begin{equation}
\{\tilde A,\tilde B \}=\tilde {\{A,B\}_{DB}}~~;~~
\{\tilde A,\tilde B \}\mid _{\phi =0}=\{A,B \}
 _{DB}~~;~~\tilde 0=0,
\label{b11}
\end{equation}
plays a crucial role in this scheme in making contact with the DBs.
Hence the outcome
of the BT extension is the closed system of FCCs with the FC
Hamiltonian given below,
\begin{equation}
\{\tilde \Theta _\alpha ^\mu ,\tilde \Theta _\beta ^\nu\}=
\{\tilde \Theta _\alpha ^\mu ,\tilde \Psi _l\}=
\{\tilde \Theta _\alpha ^\mu ,\tilde H\}= 0~~;~~
\{\tilde \Psi _l ,\tilde \Psi _n\}\approx 0 ~;~
\{\tilde \Psi _l ,\tilde H\}\approx 0.
\label{b12}
\end{equation}
We will see that due to the non-linearity in the SCCs, the extensions
 in the improved variables, (and subsequently in the FCCs
 and FC Hamiltonian),
turn out to be infinite series. This type of situation has been encountered
 before \cite{bbg,sg1}. In the present case, the solution for $X^{\mu\nu}_
 {\alpha\beta}$ in (\ref{b2}) is obtained as,
\begin{equation}
X^{\mu\nu}_{\alpha\beta}=
\left (
\begin{array}{cc}
-\bar\Delta _{11}^{\mu\nu} & \frac {1}{2}g ^{\mu\nu}\\
2\bar\Delta _{12}^{\mu\nu} & 0
\end{array}
\right ) .
\label{b13}
\end{equation}
The inverse of the above matrix, as defined in  (\ref{b6}), is
\begin{equation}
X_{\mu\nu}^{\alpha\beta}=
\left (
\begin{array}{cc}
0 & \frac {1}{2}(a_1P_\mu P_\nu +a_2g_{\mu\nu})\\
2g_{\mu\nu} & a_2(\bar\Delta _{11})_{\mu\nu}
\end{array}
\right ) .
\label{b14}
\end{equation}
The parameters $a_1$ and $a_2$ have already been defined in (\ref{21}).
Using (\ref{b2}), the one-$\phi $ BT extensions in the SCCs are,
$$
\tilde \Theta _1 ^{\mu (1)} = X^{\mu\nu}_{11}\phi^1_\nu +
X^{\mu\nu}_{12}\phi^2_\nu =-\bar\Delta _{11}^{\mu\nu}\phi^1_\nu
+\frac{1}{2}\phi^{2\mu } ~,$$
\begin{equation}
\tilde \Theta _2 ^{\mu (1)} =
X^{\mu\nu}_{21}\phi^1_\nu+X^{\mu\nu}_{22}\phi^2_\nu   =
2\bar\Delta _{12}^{\mu\nu} \phi^1_\nu .
\label{b014}
\end{equation}
It is easy to convince oneself that the $B^{\mu\nu}_{\alpha\beta}$
functions defined in (\ref
{b3},\ref{b4},\ref{b5}) are in general non-vanishing giving rise to terms
having higher powers in BT variables. Let us now compute
the one-$\phi $ extension in the $\tilde x_\mu $, {\it i.e.} the
improved variable corresponding to $x_\mu $, the canonical
coordinate variable. Simplifying (\ref{b7}), we get,
$$
\tilde x_\mu ^{(1)}=-(\phi ^\alpha )^\nu\epsilon _{\alpha\beta }
X^{\beta\gamma}_{\nu\sigma} \{\bar \Theta ^\sigma _\gamma ,x_\mu\}
_{(q)} $$
$$=[2(S_{\nu\mu}+k_1\pi g_{\nu\mu})+\frac{1}{M}(k_2\phi -1)a_2
(\bar\Delta _{11})_{\nu\mu}](\phi ^1)^\nu  $$
\begin{equation}
+\frac{1}{2M}(1-k_2\phi )(a_1P_\mu P_\nu +a_2g_{\mu\nu})(\phi ^2)^\nu  .
\label{b15}
\end{equation}
Hence, up to one-$\phi ^\alpha _\mu $ BT extension,  it is
straightforward to check the following relation,
\begin{equation}
\{x^\mu +\tilde x^{(1)\mu },x^\nu +\tilde x^{(1)\nu}\}=-\frac{S^{\mu\nu}}{M^2}
+(\phi^\alpha -terms)~.
\label{b16}
\end{equation}
In the above calculation, one has to remember that the BT extended
expressions for the FCCs has to be used.
The PB between the {\it full} $\tilde x_\mu$, ({\it i.e.} to all
orders in $\phi ^\alpha $), will satisfy
\begin{equation}
\{\tilde x^\mu ,\tilde x^\nu\}=
-\frac{\tilde S^{\mu\nu}}{M^2}.
\label{gau}
\end{equation}

To ascertain the one-$\phi ^\alpha _\mu $
term in the right hand side of the above PB, one has to compute
extensions up to two-$\phi ^\alpha _\mu $ in $\tilde x_\mu $.
Thus, we have explicitly derived the following relation,
\begin{equation}
\tilde x_\mu =x_\mu  + \tilde x_\mu ^{(1)}+(higher~\phi ^\alpha -terms) ~,
\label{b17}
\end{equation}
with $\tilde x_\mu ^{(1)}$ given by (\ref{b15}). This is the
cherished mapping between the NC space-time coordinate $\tilde
x_\mu$ and the usual space-time coordinate $x_\mu$. A similar
type of mapping between an NC coordinate and a canonical
coordinate was also proposed, (in a non-relativistic framework),
in \cite{cnp}. However, it should be pointed out that $x_\mu$ in
the above mapping (\ref{b17}) is truly the usual space-time coordinate, with
correct Lorentz transformation properties, whereas the canonical
coordinate introduced in \cite{cnp} is not. The BRST quantization of this irreducible SCC system does not pose any technical complications.

The presence of non-commutativity has made a strong impact in recent years in High Energy Physics, ever since the appearance of the seminal work of Seiberg and Witten \cite{sw}. Noncommutativity is induced in the open string boundaries, when the string moves in a constant two-form background Neveu-Schwarz field (or equivalently a magnetic field). The noncommutativity parameter is identified with the inverse of the constant magnetic field.  Recently we have shown \cite {bcg} that contrary to previous works \cite{jab} involving constraints, the non-trivial mixed boundary conditions are responsible for this noncommutativity. However, the issue of noncommutativity is quite alive and different avenues have to be explored to obtain the above feature. Exploiting the observation that noncommutativity appears in a particular choice of regularization in string theory, Seiberg and Witten \cite{sw} have provided an explicit map, (to the lowest nontrivial order in the noncommutativity pa!
rameter), connecting noncommutative and ordinary gauge fields.The idea of equivalence between gauge orbits in ordinary and noncommutative space-time plays a pivotal role in extablishing the map. Apparantly, the noncommutativity of space-time is not exploited directly since one stays in ordinary the space-time and introduces extra interaction  terms in the original model as noncommutative effects.

On the other hand, in \cite{sgnew} we have shown a new way of interpreting the Seiberg-Witten map \cite{sw} which is more geometric in nature and rests directly on space-time noncommutativity. In \cite{sgnew} the Seiberg-Witten map appears as one changes the argument of the gauge field from ordinary to noncommutative space-time in a particular  way. This means that without going in to the concept of identifying gauge equivalence between noncommutative and commutative space-times, it is possible to recover the Seiberg-Witten map in a geometrical way. The spinning particle lives in an extended space having noncommutative and commutative sectors and gauge fields in these sectors can be connected by the Seiberg-Witten map.
Precisely in this context the above spinning particle model can become relevant since they provide a natural framework for introducing noncommutativity in space-time, without any need to bring in external interactions. The details of this mechanism will be reported elsewhere.
\vskip .2cm
\begin{center}
{\bf V. PROJECTION OPERATOR METHOD}
\end{center}
\vskip .1cm
Recently a new scheme, the Projection Operator Method, has been
proposed by Batalin, Lyakhovich and Marnelius \cite{blm}, where
one is able to quantize a constrained system, having a set of
reducible SCCs and FCCs, in a manifestly covariant framework.
A generalized BRST operator has also been posited in \cite{blm}.
The formalism is in complete contrast to the Auxiliary Variable
approach \cite{bn} and BT extension \cite{bt} discussed in the
previous sections. Here no non-physical degrees of freedom are
introduced. Instead, the major task is the construction
of an invariant projection operator that projects out the
maximal subset of constraints in involution, ({\it i.e.}
FCCs), from the full (reducible) set of constraints . With this set of
reducible FCCs one can attempt a
BRST quantization. However, the presence of SCCs causes an obstruction,
which requires a generalization of the BRST operator \cite{blm}. In
\cite{blm} the authors make the crucial assumption that
for a (reducible) set of constraints, with the PB algebra,
\begin{equation}
\{\Theta _\alpha,\Theta _\beta\}=U_{\alpha\beta}^{~~\gamma}
\Theta _\gamma+\Delta _{\alpha\beta},
\label{b18}
\end{equation}
one can construct a suitable projection matrix $P_\alpha^{~\beta}$
satisfying,
\begin{equation}
P_\alpha^{~\beta}\Delta _{\beta\gamma}P_\chi ^{~\gamma}=0~~,~~
P_\alpha ^{~\beta}P_\beta^{~\gamma}=P _\alpha^{~\gamma}.
\label{b19}
\end{equation}
This will project out  the reducible set of FCCs ,
\begin{equation}
\Theta _\alpha'=P_\alpha ^{~\beta}\Theta _\beta~~,
~~\{\Theta _\alpha',\Theta _\beta '\}\approx 0.
\label{b20}
\end{equation}
It is imperative to show that the above assumption works in a
non-trivial model and the present work is probably the first one
where its validity is demonstrated explicitly. The formalism
\cite{blm} is applicable even in systems where one can not
separate out the FCCs and SCCs without spoiling manifest
covariance. However, as shown in section {\bf II}, in our model
this separation is possible. This slightly simplifies the problem
since we have to consider only the reducible SCCs $\Theta _\alpha
^\mu $ in (\ref{12}).

The all important projection operator is given by the
following matrix,
\begin{equation}
P_\alpha ^{~\beta}=
\left (
\begin{array}{cc}
\frac{1}{M^2}S^{\mu\sigma}P_\sigma P^\nu & S ^{\mu\nu}\\
\frac{1}{M^2}P^\mu P^\nu & (g^{\mu\nu}
+\frac{1}{M^2}P^\mu S^{\nu\sigma}P_\sigma )
\end{array}
\right ) .
\label{b21}
\end{equation}
This will lead to the FCCs $\Theta _\alpha'=
P_\alpha ^{~\beta}\Theta _\beta$ whose closure property can be
directly checked. Now one must enlarge the phase space by introducing ghosts, ghosts for ghosts etc. We follow the prescription of \cite{blm} and for a generic $L$'th stage reducible theory introduce  the ghosts ${\cal C} ^a ,\bar{\cal P} _a $ and secondary ghosts
${\cal C}'^{a_{r}},\bar {\cal P}'_{a_{r}}$ where $r=1,..,L$ and the ghosts satisfy
$$
\epsilon ({\cal C} ^a)=\epsilon (\bar {\cal P}_a)=\epsilon_a +1~,~
\epsilon ({\cal C}'^{a _{r}})=\epsilon (\bar {\cal P}'_{a _{r}})=\epsilon_{a _{r}}+r +1~,~r=1,..,L,$$
$$
\{{\cal C} ^a ,\bar {\cal P}_b\}=\delta ^a _b~,~\{{\cal C}'^{a _r} ,\bar {\cal P}'_{b _r}\}=\delta ^{a _r} _{b _r}~,~r=1,..,L,$$
\begin{equation}
gh({\cal C} ^a)=-gh(\bar {\cal P}_a )=1~,~gh({\cal C}'^{a _r})=-gh(\bar {\cal P}'_{a _r} )=r+1 .
\label{gho}\end{equation}
Here $\epsilon ({\cal A})$ and $gh({\cal A})$ denote the parity and ghost number of ${\cal A}$ and in the above $\epsilon (T_a)\equiv \epsilon _a$, where $T_a$ represents the full set of constraints. Let us now construct the following odd real function $\Omega $ with ghost number $1$,
$$
\Omega ={\cal C} ^a T_a +{\cal C}'^{a _1 }Z^a_{a_1}\bar {\cal P}_a (-1)^{\epsilon _a}+\sum ^L_{r=2}{\cal C}'^{a _r }Z^{a _{r-1}}_{a _r}\bar {\cal P}_{a _{r-1}} (-1)^{\epsilon _{a _{r-1}}}
$$
$$
+(-1)^{\epsilon _b}\frac{1}{2}{\cal C} ^b {\cal C} ^a \Delta _{ab}^e \bar {\cal P}_e (-1)^{\epsilon _e} +
(-1)^{\epsilon _a}{\cal C} ^a {\cal C'} ^{b _1} \Delta _{b _1a}^{a _1} \bar {\cal P'}_{a _1}(-1)^{\epsilon_ {a_1}}
$$
\begin{equation}
+(-1)^{(\epsilon_b +\epsilon _a \epsilon_e )}\frac{1}{6}{\cal C} ^e {\cal C} ^b {\cal C} ^a  \Delta _{a be}^{a _1} \bar {\cal P'}_{a _1}(-1)^{\epsilon_ {a_1}}
+...
\label{blm1}
\end{equation}
In the above we have defined the constraint algebra as,
$$\{T_a,T_b\}=C_{ab}+\Delta _{ab}^eT_e,$$
and the reducibility conditions as,
$$Z^a_{a_r}T_a=0~,~r=1,..L.$$
The higher order reducibility conditions are of the form,
$$Z_{a_2}^{a_1}Z_{a_1}^a=0,$$
and so on.
The higher order structure function $\Delta _{b _ra}^{a _r}$ follows from the requirement $$\{T_a, Z_{a_r}^bT_b\}=0,$$ wheras $\Delta _{a be}^{a _1}$ is induced by the Jacobi identity $$J(T_a,T_b,T_e)\equiv \{\{T_a,T_b\},T_e\}+~cyclic~ terms=0.$$ Rest of the structure functions will follow from still higher order Poisson Brackets. For a detailed discussion see for example \cite{hen}.

 In the present case, $T_a\equiv (\Psi _1,\Psi _2,\Theta ^\mu _1,\Theta ^\nu _2)$ in (\ref{11},\ref{12}) and our model is single stage reducible with $L=1$ (\ref{15}). One can ascertain that  $\Delta _{b_1a}^{a _1}=0$ from $\{T_a, P^\mu\Theta _{1\mu }\}=0$. $\Delta _{abe}^{a _1}$ contributes only from the non-trivial Jacobi identity expression,
 \begin{equation}
 J(\Theta _1^\mu ,\Theta _1^\nu,\Theta _2^\sigma )=\frac{1}{M}(g^{\sigma\nu}P^\mu-g^{\sigma\mu}P^\nu)\Psi _1 +(g^{\sigma\nu}\Theta_2^\mu-g^{\sigma\mu}\Theta _2^\nu)\Psi _1.
\label{blm2}
\end{equation}
We drop the second term since it is quadratic in the constraints and hence is strongly vanishing. The rest of the higher order structure functions are zero.
With the above inputs, the function $\Omega$ is
$$
\Omega={\cal C} ^1\Psi _1+{\cal C} ^2\Psi _2+{\cal C} ^{1\mu}\Theta _{1\mu}+{\cal C} ^{2\mu}\Theta _{2\mu}
$$
$$+{\cal C}'P^\mu\bar {\cal P}_{1\mu}+\frac{1}{2}{\cal C} ^{2\mu}{\cal C} ^{1\nu}((g_{\mu\nu}P_\sigma-g_{\mu\sigma}P_\nu)\bar {\cal P}_2^\sigma +\frac{1}{M}g_{\mu\nu}\bar {\cal P}_1))+\frac{1}{2}{\cal C} ^{1\mu}{\cal C} ^{1\nu}S_{\mu\nu}\bar {\cal P}_1
$$
\begin{equation}
+{\cal C} ^{1\mu}P_\mu{\cal C} ^{1\nu}\bar {\cal P}_{1\nu}+\frac{1}{6M}{\cal C} ^{2\mu}{\cal C} _{1\mu}{\cal C} ^{1\nu}P_\nu\bar {\cal P}'~.
\label{blm3}
\end{equation}
Next an even real function $\Pi$ with ghost number zero is posited to be
\begin{equation}
\Pi ={\cal C} ^a P^b _a \bar {\cal P}_b (-1)^{\epsilon_b}+\sum_{r=1}^L(-1)^r{\cal C} ^{a_r}P'^{b_r}_{a_r}\bar {\cal P}_{b_r}(-1)^{\epsilon_{b_r}}+...
\label{blm4}
\end{equation}
where the dots indicate higher powers of ghost terms.
The functions $P'^{b_r}_{a_r}$ are yet to be determined from the following relations \cite{blm},
\begin{equation}
\{\Pi,\{\Pi, \Omega\}\}=\{\Pi, \Omega\}~,~\{\Pi,\{\Pi, \{\Omega ,\Omega \}\}\}=\{\Pi, \{\Omega ,\Omega\}\}.
\label{blm5}
\end{equation}
This leads to the cherished form of the generalized BRST charge
\begin{equation}
\Omega '\equiv\{\Pi ,\Omega \}
\label{blm6}
\end{equation}
which obeys the nilpotency property,
\begin{equation}
\{\Omega ',\Omega '\}=0.
\label{blm7}
\end{equation}
In the present case, using (\ref{b21}), we get,
$$
\Pi=
{\cal C}^1\bar{\cal P}_1+{\cal C}^2\bar{\cal P}_2+
\frac{1}{M^2}{\cal C}^{1\mu}S_{\mu\sigma}P^\sigma P^\nu\bar{\cal P}_{1\nu}
$$
\begin{equation}
+{\cal C}^{1\mu}S_{\mu\nu}\bar{\cal P}^\nu_2 +\frac{1}{M^2}{\cal C}^{2\mu}P_\mu P^\nu\bar{\cal P}_{1\nu}+\frac{1}{M^2}{\cal C}^{2\mu}P_\mu S^{\nu\sigma}P_\sigma\bar{\cal P}_{2\nu}-{\cal C}'P'\bar{\cal P}'.
\label{pi}
\end{equation}
Computation of the last term is straightforward but tedious and is not pursued here. Just as in the irreducible case, in general one has to modify the Hamiltonian so that it has vanishing PB with the BRST charge $\Omega '$. However, in the present case the Hamiltonian remains the same as in (\ref{ham}) due to its simple structure. In the subsequent quantization these functions are elevated to quantum operators and obviously they have to be properly ordered.

\vskip .2cm
\begin{center}
{\bf VI. CONCLUSIONS}
\end{center}
\vskip .1cm The covariant quantization of the Nambu-Goto spinning
particle model is analyzed and the relevance of this model in
inducing a non-commutative space-time is demonstrated. The
technical problem of covariant quantization in the present model
is very subtle since the set of constraints comprise {\it
reducible Second Class Constraints} apart from First Class
Constraints. (The above classification follows from the
prescription of Dirac \cite{d}.) Special methods have been devised to
tackle the above mentioned reducibility problem. We have discussed
here two schemes: (i) the auxiliary variable method \cite{bn} where the
phase space is enlarged in an appropriate way and (ii) the
projection operator method \cite{blm}, where a (reducible) set of first class
constraints is projected out from the set of second class
constraints. The latter formalism has been proposed very recently
and the present work constitutes a non-trivial application of the
same. Construction  of the projection operator as well as the necessary steps for the BRST quantization  has been provided.

A number of projects to be pursued further immediately comes to mind: Firstly
a thorough appraisal of the mapping between non-commuting and
ordinary space-time that has been exhibited here,
in the light of \cite{sgnew}, and secondly
a quantum BRST analysis of the model in the projection
operator formalism taking into account the operator ordering problems. Also it has been suggested \cite{rj} that the
Faddeev-Jackiw \cite{fj} method of  symplectic quantization may be useful in the context
of covariant quantization of the spinning particle model. Work is in progress in
this direction as well.
\vskip 1cm
\noindent
{\bf Acknowledgements}: It is a pleasure to thank Professor R.Banerjee and Professor R.Jackiw
for helpful comments.

\end{document}